\documentstyle[12pt]{article} 
\def\be{\begin{equation}}
\def\ee{\end{equation}} 
\def\bea{\begin{eqnarray}}
\def\eea{\end{eqnarray}} 
\def\ba{\begin{array}} 
\def\ea{\end{array}}
\def\pa{\partial}

\def\ga{\gamma} 
\def\Ga{\Gamma} 
\def\si{\sigma}
 
\def\de{\delta} 
 
\def\al{\alpha}
\def\vphi{\varphi} 
 
\def\la{\lambda} 
\def\La{\Lambda}
\def\eps{\epsilon}

\def\dk{\delta k} 
\def\ln{{\rm ln}}
\def\Tr{{\rm Tr}} 
 
\def\nn{\nonumber} 

\begin{document}
\begin{center}

{\LARGE The Price of an Exact, Gauge-Invariant RG-Flow Equation}\\

\vspace*{1cm}
Vincenzo Branchina\footnote{Vincenzo.Branchina@ires.in2p3.fr}$^{a,b}$,
Krzysztof A. Meissner\footnote{Krzysztof.Meissner@fuw.edu.pl}$^{c}$,
Gabriele Veneziano\footnote{Gabriele.Veneziano@cern.ch}$^{a}$\\
\vspace*{1cm}
$^a$  Theory Division, CERN, CH-1211 Geneva 23, Switzerland \\
$^b$ IReS, Theoretical Physics Group, ULP and
CNRS\\ 23 rue du Loess, 67037 Strasbourg, France\\ 
$^c$ Institute of Theoretical Physics, Warsaw University\\ Ho\.za 69, 
00-681 Warsaw, Poland \\

\vspace*{1cm}

{\LARGE Abstract}\\
\end{center}

We combine old ideas about exact renormalization-group-flow (RGF) 
equations with the
Vilkovisky-De Witt (VDW) approach to reparametrization invariant
effective actions and arrive at a new, exact, gauge-invariant
RGF equation. The price to be paid for such a result is that both the 
action and the RGF equation depend explicitly upon the base point 
(in field space) needed for  the VDW construction. We briefly discuss 
the complications originating from this fact and possible ways to 
overcome them.

\vspace*{1.5cm}

The idea of renormalization, originally introduced to remove
infinities from perturbative calculations, has evolved into a
powerful tool that helps  understanding the global behaviour of
quantum and statistical systems under changes of the
observation scale\cite{kw,sw}. 

The search for new, non-perturbative methods to handle problems out
of the reach of perturbation theory has prompted in recent years a
renewed and growing interest \cite{pol,has,wett,mar,mor} in the
``old" subject \cite{kw,wh} of ``exact" renormalization group (RG)
equations. One typically defines a scale-dependent effective
action, $\Ga_k$, which interpolates between the classical (bare)
action $S$ at $k=\La$ (the UV cutoff) and the effective action
$\Ga$ at $k=0$. The free term is 
modified by the introduction of a suitable (but largely 
arbitrary) cutoff function that effectively kills
the contribution to the functional integral from momenta below the
running scale $k$. 
The implementation of such a procedure for gauge theories 
poses however a major problem: the presence of the cutoff function 
prevents the possibility of defining a gauge invariant $\Ga_k$
(see \cite{rw} for earlier attempts to circumvent this problem).

In this letter we follow the spirit of Ref. \cite{kw,wh}, that 
of direct integration over successive shells of degrees 
of freedom, and combine this idea with the 
geometrical approach pioneered by Vilkovisky and De Witt \cite{Vilko,DeWitt} 
(see also \cite{FT}) in order to define a gauge-invariant (more generally 
a reparametrization-invariant)  effective action. 
The final outocome will be an exact, gauge-invariant, RGF equation that, to 
the best of our knowledge, has never appeared before in the literature. It 
contains an explicit dependence upon the base-point (in field space) that 
enters the VDW construction, a price that we believe to be inevitable for 
achieving the goal. We shall comment at the end on the possible complications
due to such dependence when one applies it to specific problems.
As for the derivation of the equation itself, it will be oulined, for 
pedagogical reasons, in three steps. We first  present
a new (and in our opinion more transparent) derivation of basically 
known results for scalar theories. The main new results follow as we 
turn to the case of reparametrization-invariant effective actions a 
la VDW and their exact RGF equation. The final step, going over to the 
case of gauge theories, is then  straightforward, as often emphasized  
by Vilkovisky.

We thus begin by defining $\Ga_k$ for a simple scalar field theory. 
If $\La$ is the UV cutoff, we introduce the notation $\phi_{0}^{\La}$ 
for the field, to indicate that it contains ``modes"  in the range 
$[0,\La]$, and write the classical (bare) action as $S[\phi_{0}^{\La}]$. 
For any given scale $k$, we divide  
$\phi_{0}^{\La}$ into the ``low-frequency" and ``high-frequency" components,
$\phi_{0}^{k}$ and $\phi_{k}^{\La}$ respectively, where $\phi_{0}^{k}$
contains the modes $\phi_p$ with $0<p<k$, and $\phi_{k}^{\La}$ those 
in the range $[k,\La]$. Even though for the scalar theory 
it is always possible to define the RG flow in Fourier space, it is well 
known that the notion of RG flow is much more general. Neither 
$k$ nor $\La$ must necessarily have the meaning of momenta (this observation 
is important for the following where we have to implement a gauge invariant 
flow for gauge theories).    

Let us now introduce the notion of ``shell", 
described by $\de k$, denote the fields $\phi_0^{k-\delta k}$, 
$\phi_{k-\de k}^k$ and $\phi_k^\Lambda$ by $\phi_{{}_{<}}$, 
$\phi_{{}_{S}}$ and $\phi_{{}_{>}}$, respectively, and use 
de Witt's \cite{DeWitt} condensed notation whereby an index such as $i$
denotes all indices (Fourier, Lorentz, spinor, space-time 
coordinate $x$, \dots). Repeated indices will denote summation 
over  internal indices as well as integration over space-time (or momenta). 
The components of $\phi_{{}_{S}}$ and  $\phi_{{}_{>}}$ 
will be indicated by $\phi^s$ and $\phi^a$ (same for $\bar{\phi})$, 
and differentiation w.r.t. any $\phi^i$ ($\bar{\phi}^i$) by a 
comma followed by the index $i$. Later on we will also use 
$A,B,\ldots$ to denote fields with components in the slightly
larger interval $[k-\de k,\La]$.

The effective action $\Ga[\bar{\phi}]$, a functional of the ``classical" 
(or ``mean") fields $\bar{\phi}$, can be defined as the solution of 
the functional-integral equation:
\be\label{Gamma}
e^{-\Ga[\bar{\phi}]} = \int[D{\phi}] e^{- S[{\phi}] +
({\phi}^i - \bar{\phi}^i )\Ga[\bar{\phi}],_i} \, .
\ee

The scale-dependent generalization of (\ref{Gamma}) that we propose to
use, and later generalize, is simply
obtained from  (\ref{Gamma}) after inserting under the integral a product
of $\delta$-functions, $\Pi_0^k \delta({\phi}_p - \bar{\phi}_p)$, i.e.:

\be\label{gammak}
e^{-\Ga_k[\bar{\phi}]} =
\int[D{\phi}_{{}_{>}}] e^{- S[\bar{\phi}_{{}_{<}}, {\phi}_{{}_{>}}] +
({\phi}^a - \bar{\phi}^a ) \Ga_k
[\bar{\phi}],_a} \, .
\ee
This very natural definition of a scale-dependent effective 
action clearly interpolates between the classical and the quantum action, 
$\Ga_{\La}[\bar{\phi}] = S[\bar{\phi}]$ and
$\Ga_0[\bar{\phi}]=\Ga[\bar{\phi}]$, and can be obtained by
a partial Legendre transform  \cite{Zumino} of a functional 
$W_k[\bar{\phi}_0^k, J_k^{\La}]$ in which the low-frequency fields 
$\bar{\phi}_0^k$ are kept as parameters, while the high frequency degrees
of freedom are Legendre-transformed.

We now derive some identities that will be useful in the following.  By
differentiating $\Ga_k$ in Eq.(\ref{gammak}) w.r.t. $\bar{\phi}^a$, we find
(for a non-singular 2nd-derivative matrix of $\Ga_k$)
\be\label{expe}
<\phi^a > =  \bar{\phi}^a \ ,
\ee
where the average is computed with the weight in Eq. (\ref{gammak}). 
Thus, as we expect,
$\bar{\phi}^a$ is the mean value of $\phi^a$.  
Differentiating Eq.(\ref{expe}) w.r.t. $\bar{\phi}^s$ we get:
\bea\label{rel1}
&<&S(\bar{\phi}_{{}_{<}},\bar{\phi}_{{}_{S}},\phi_{{}_{>}} )_{, s}
(\phi^a- \bar{\phi}^a)> \nn\\ 
&=&\Ga_{k ,\, s b} <(\phi^b- \bar{\phi}^b)(\phi^a- \bar{\phi}^a) > = 
\Ga_{k ,\, s b}(\Ga_{k,b a})^{-1} \; ,
\eea
where $-(\Ga_{k,b a})^{-1}$ is the propagator for 
modes above the shell. A second useful relation comes from differentiating 
$\Ga_k$ w.r.t. $\bar{\phi}^s$: 
\be\label{rel2}
<S(\bar{\phi}_{{}_{<}},\bar{\phi}_{{}_{S}},\phi_{{}_{>}})_{, s}> = \Ga_{k,\, s}
\; .  
\ee
Finally, differentiating $\Ga_k$ once more w.r.t. $\bar{\phi}^s$, 
and making use of Eq.(\ref{rel1}), we obtain the relation:  
\bea\label{rel3} 
&<&S_{, s s'}> - <S_{, s} S_{, s'}> + <S_{, s}> < S_{, s'}>  \nn \\
&=&\Ga_{k ,\, s s'} - \Ga_{k ,\, s a} (\Ga_{k,a b})^{-1}
\Ga_{k ,\, b s'} \; .
\eea

Let us consider now the effective action $\Ga_k$ at a slightly lower scale
$k-\de k$. From Eq.(\ref{gammak}) we have :
\be\label{gammadk}
e^{-\Ga_{k-\de k}[\bar{\phi}_0^\La]} =\int[D \phi_{{}_S}]
e^{(\phi^s-\bar{\phi}^s)\Ga_{k-\de k,\,s}} e^Y \; ,
\ee
where 
\be\label{aus} 
e^Y = \int[D \phi_{{}_{>}}] 
e^{- S[\bar{\phi}_{{}_{<}},\phi_{{}_S}, \phi_{{}_{>}}] +
(\phi^a - \bar{\phi}^a )\cdot \Ga_{k -\de k, a}} \, .
\ee
We are interested in computing the difference between $\Ga_{k}$ and
$\Ga_{k-\de k}$ to $O(\de k)$ and thus start expanding to first
order $\Ga_{k-\de k, a}$ around $\Ga_{k,a}$ in Eq.(\ref{gammadk}). At
the same time we expand $S[\bar{\phi}_{{}_{<}},\phi_{{}_S}, \phi_{{}_{>}}]$
around $\phi_{{}_S}=\bar{\phi}_{{}_S}$. Denoting the fluctuations
$(\phi^s-\bar{\phi}^s)$ and $(\phi^a-\bar{\phi}^a)$ by $\eta^s$ and $\eta^a$
respectively, we get :
\be\label{gammaup} 
e^Y = e^{- \Ga_k} 
<e^{-[S_{, s}\eta^s+\frac{1}{2}S_{, s s'} 
\eta^s\eta^{s'} +  \dots
+\delta k \frac{\pa \Ga_{k, a}}{\pa k}\eta^a]}> \ ,
\ee
where the (omitted) arguments of $S_{, s}$ and $S_{, s s'}$ are 
$[\bar{\phi}_{{}_{<}},\bar{\phi}_{{}_S}, \phi_{{}_{>}}]$.

Following the classic arguments of \cite{wh}, we know that, in
order to collect all terms up to $O(\delta k)$, we only need to keep 
terms up to $O((\eta^s)^2)$, and thus we neglect the ellipses. 
The r.h.s. of Eq.(\ref{gammaup}) can be now computed using the identity
\be
\left<e^{-f}\right>=e^{-<f>+\frac12(<f^2>-<f>^2)+O(f^3)}.
\label{aviden}
\ee
Thanks to (\ref{expe}), the last term 
in (\ref{gammaup}) can only contribute $O((\dk)^2)$, so we also 
neglect this term. Then, with the help of the relations (\ref{rel2}) 
and (\ref{rel3}), we immediately compute the r.h.s. of Eq.(\ref{gammaup}) 
and find that (\ref{gammadk}) becomes : 
\be\label{simple}
e^{-\Ga_{k-\de k}} =e^{- \Ga_k}\int[D \eta_{{}_S}]
e^{\Delta\Ga_{k,\,s}\eta^s 
-\frac{1}{2} K_{ s s'} \eta^s\eta^{s'}}  \, ,
\ee
where $\Delta\Ga_{k,\,s}= \Ga_{k-\de k ,\,s} - \Ga_{k,\,s}$ and 
$K_{ s s'}$ is nothing but the r.h.s. of Eq.(\ref{rel3}), i.e. :
\be\label{K's}
K_{ s s'} = \Gamma_{k ,\, s s'} - \Gamma_{k ,\, s a} (\Gamma_{k,a b})^{-1}
\Gamma_{k ,\, b s'} \; .
\ee

As $\Delta\Ga_{k,\,s}$ is $O(\dk)$, it would contribute an $O((\dk)^2)$ term
after performing the gaussian 
integral. Neglecting 
again this higher order term, we finally find that the difference 
between $\Ga_{k - \de k}$ and $\Ga_k$  (evaluated at the same values 
of their arguments) consists, to $O(\de k)$, of just a determinant, i.e. 
\be 
\label{final} 
\Ga_{k -\de k} = \Ga_k + \frac12 \Tr\, \ln\, K_{s s'} . 
\ee
Using standard properties of determinants, Eq. (\ref{final}) can be rewritten
in a form that will  be more useful for our subsequent generalizations, i.e.
\be 
\label{finaldet} 
\Ga_{k -\de k} - \Ga_k  = \frac12  
\ln \, \left( \frac{{\rm det} \Gamma_{k,AB}}
{{\rm det} \Gamma_{k,a b}} \right) \  , 
\ee
where we recall that the capital indices $(A, B)$ span the region
$[k-\de k,\La]$, while $(a, b)$ are for the region $[k,\La]$.

Eq.(\ref{final}) was already derived in \cite{nico} for the case of 
a spin hamiltonian $H(\si_{\bf p})$ (where $\si_{\bf p}$ is the 
Fourier component of the spin field)
following a different, though equivalent, line of reasoning. The derivation 
presented above is new and, furthermore, is more suitable 
for extension to the more general cases we shall consider 
below. This is why we have presented the different steps 
in great detail\footnote{Note also 
that a rederivation of Eq. (\ref{finaldet}), equation that already
appeared in a previous version of the present paper, was given in
\cite{correia}.}.

Let us now discuss how one can extend our results to the general case,
including gauge theories.  It was first noted by Vilkovisky \cite{Vilko}
that the usual definition of the effective action, Eq.(\ref{Gamma}), is 
in general not invariant under a reparametrization of the classical fields. 
Obviously this holds true also for our definition (\ref{gammak}) of $\Ga_k$ 
at any scale $k$. He also pointed out that, in the case of gauge theories, 
the gauge dependence of the off-shell effective action is 
just a manifestation of such a reparametrization dependence.

The origin of the problem can be seen easily from the definition
of the effective action (\ref{Gamma}). Let us think of the 
(field) configuration space as a manifold  ${\cal M}$ endowed with a metric $g_{ij}$
and assume that $\Ga$, like $S$, is a
scalar field on  ${\cal M}$. While the functional integration
measure can be made reparametrization invariant through the introduction
of a $\sqrt{g}$, the second term in the exponential has bad transformation
properties since the gradient is a covariant vector while the 
``coordinate difference" ($\phi - \bar{\phi}$) is a contravariant vector 
only if the $\phi$'s are euclidean coordinates in a 
trivial (flat) manifold. In the case of gauge theories there is an additional 
complication coming from the fact that the physical space is the quotient space 
${\cal M}/{\cal G}$ (${\cal G}$ is the 
gauge group) rather than ${\cal M}$. We'll came back on this point later.

Vilkovisky and De Witt have shown that a meaningful 
definition of the effective action can be given also in the general (curved) case
in terms of {geodesic normal fields}, $\si^i[\vphi_*,\phi]$, based at a point 
$\vphi_*$ in ${\cal M}$\cite{Vilko,DeWitt}.
The $\si^i[\vphi_*,\phi]$ are the components of a vector tangent at
$\vphi_*$ to the geodesic from $\vphi_*$ to $\phi$. Its length
is the distance between the two points along the geodesic itself.
Under coordinate transformations $\si^i[\vphi_*,\phi]$
 transforms as a vector at $\vphi_*$ and as a scalar at
$\phi$. A useful property of the $\si$ fields is that any scalar 
function $A[\phi]$ can be expanded in a covariant 
Taylor series\cite{Vilko,DeWitt} (the semicolon denotes covariant 
derivatives w.r.t. $\phi$) : 
\be\label{covaga}
A[\phi] \equiv A[\vphi_*,\si]= \sum_{n=0}^{\infty}\frac{1}{n!}
A_{\,;\, a_1 \cdots\, a_n}[\vphi_*]
\si^{a_1} \cdots \si^{a_n}  \, .
\ee
As emphasized before, the definitions of the upper space, of the shell, 
and of the lower space are completely general and can be obtained with 
the help of any mode decomposition of the fields. From now on we denote 
by $\lambda$ these generic modes. 
As before we introduce the notation $\si^i = 
(\si_{_{<}}, \si_{_{S}}, \si_{_{>}})$. The submanifold spanned by $\si_{_{>}}$ 
we denote by  ${\cal M_{_{>}}}$ and the one spanned by $(\si_{_{S}}, \si_{_{>}})$
by ${\cal M_{_{\ge}}}$.
The metric in $\si$ coordinates is related to the original metric by
\be\label{inducedm}
\hat g_{_{lm}}(\vphi_*,\si)=\frac{\partial \phi^i}{\partial \si^l}
\frac{\partial\phi^j}{ \partial \si^m}\,g_{_{ij}}(\phi)\ .
\ee
The induced metric on ${\cal M_{_{>}}}$ (${\cal M_{_{\ge}}}$) 
is just the restriction of $\hat g_{_{lm}}$ to the appropriate set of 
indices, $\hat g_{_{ab}}$ ($\hat g_{_{AB}}$).

Given the arbitrary coordinates (fields) $\phi^i$, the base 
point $\vphi_*$, and the gaussian normal coordinates 
$\si^{i}$ in ${\cal M}$, we can now define, following \cite{DeWitt}, 
the scale (i.e. $\lambda$)-dependent effective action, 
$\hat\Ga_\lambda$, as :
\be\label{gamkhat} 
e^{-{\hat\Ga}_\lambda[\vphi_*, \bar\si]} =
\int[D{\si}_{{}_{>}}] \sqrt{\hat g}\ e^{- S+ ({\si}^a - \bar\si^a)
{\hat\Ga}_\lambda[\vphi_*, \bar\si]_{, a}} \, ,  
\ee 
where $\hat g=\det\hat g_{_{ab}}$. 
$S$ is the classical action expanded as in (\ref{covaga}), where, as 
in the analogous Eq.(\ref{gammak}), the $\si_{_{<}}$ are replaced by 
the mean values $\bar\si_{_{<}}$ : 
$S=S[\vphi_*;\bar\si_{_{<}}, \si_{_{S}}, \si_{_{>}} ]$. 
Since $\vphi_*$ is kept fixed, the steps that lead from Eq.(\ref{gammak})
to the RG equation (\ref{finaldet}) can be now repeated with almost no
changes. The only modification is due to the presence 
in Eq.(\ref{gamkhat}) of the non-trivial metric factor $\sqrt{\hat g}$
(compare with Eq.(\ref{gammak}) where the metric is trivial). The
impact of this term can be easily seen from the r.h.s. of 
Eq.(\ref{gammaup}), where it contributes the additional $O(\delta k)$
term: 
\be \label{corr}
-\frac12({\rm ln}~{\rm det} \hat g_{_{AB}} - {\rm ln}~{\rm det}
\hat g_{_{ab}})=\frac12({\rm ln}~{\rm det} \hat g^{{AB}} - {\rm ln}~{\rm det}
\hat g^{{ab}})  \\ 
\ee
The final result is then:

\be \label{finalhat} 
{\hat\Ga}_{_{\lambda - \de \lambda}}[\vphi_*, \bar\si] = 
{\hat\Ga}_{_{\lambda}}[\vphi_*, \bar\si]  + \frac12
\ln ~ \left( \frac{{\rm det} \hat\Gamma_{\lambda,A}{}^{B}}
{{\rm det} \hat\Gamma_{\lambda,{a}}{}^b} \right) \, ,
\ee
where the indices are raised with the help of the corresponding
induced metrics on each submanifold that appear in 
Eq.(\ref{corr})\footnote{To be precise in
Eq. (\ref{finalhat}) the determinants of the metrics appear 
under an expectation value sign rather than being computed
at the expectation value of the field. 
We expect the difference to be insignificant.}.

We now wish to rewrite Eq.(\ref{finalhat}) in general coordinates.
Define:
\be \label{defGamma} 
{\Ga}_{_{\lambda}}[\vphi_*, \bar{\phi}]=
{\hat{\Ga}}_{_{\lambda}}[\vphi_*,\si(\vphi_*, \bar\phi)]  =
 {\hat{\Ga}}_{_{\lambda}}[\vphi_*,\bar\si]\ .
\ee
It is rather straightforward, though tedious, to connect
the partial derivatives of $\hat{\Ga}$ with respect to the
$\bar\si$'s to the partial {\it covariant} derivatives of $\Ga$ with respect to
the $\bar\phi$'s (both
taken, of course, at fixed $\vphi_*$). Consider first
these relations at the level of the full effective actions $\hat{\Ga}$ and 
$\Ga$. 

For the first derivatives the result is simply:
\be \label{rel1stder} 
{\Ga}_{,i}  =  D_i^k {\hat{\Ga}}_{,k} \ ,
\ee
where, following \cite{DeWitt}, we have introduced:
\be \label{Dik} 
  D_i^k = \frac{\pa \bar\si^k}{\pa \bar\phi^i} \;.
\ee
The bi-vector $D_i^k$ has the property that,
once contracted with a covariant vector at $\vphi_*$, converts it
into a covariant vector at $\bar\phi$, as exemplified indeed in (\ref{rel1stder}).

The relation connecting second derivatives can be put in the form:
\be \label{rel2ndder}  
   {\hat{\Ga}},_{kl} = (D^{-1})_k^i  (D^{-1})_l^j 
 \bar{\Ga}_{ij} \ ,
\ee
where
\be \label{barGa} 
 \bar{\Ga}_{ij} \equiv {\Ga}_{;ij} - \si^l_{;i j}
(D^{-1})_l^k \Ga_{,k} 
\ee
is a  second-rank tensor at $\bar\phi$. The quantity $\si^l_{; i j}$
has a  covariant expansion \cite{DeWitt} 
in the distance between $\vphi_*$ and $\bar\phi$.

The above formulae can be easily generalized to the case in which
the derivatives are restricted to lie on the ${\cal M_{_{>}}}$ 
(or ${\cal M_{_{\geq}}}$) manifold. Indeed the derivatives of 
$\hat{\Ga}_{\la}$
with respect to $\bar\si^a$ will be related to the derivatives of
$\Ga_{\la}$ with respect to generic coordinates $\xi^a$ on 
${\cal M_{_{>}}}$ by exactly the same formulae (\ref{rel1stder}), 
(\ref{rel2ndder}) where now:
\be \label{Dab} 
  D_b^a = \frac{\pa \bar\si^a}{\pa \xi^b} \;.
\ee

Using Eq. (\ref{inducedm}) we obtain our final result:
\be \label{finalGa} 
{\Ga}_{_{\la - \de \la}}[\vphi_*, \bar\phi] = 
{\Ga}_{_{\la}}[\vphi_*, \bar\phi]  + \frac12
\ln ~ \left( \frac{{\rm det} \bar{\Ga}_{\lambda\, A}{}^{B}}
{{\rm det} \bar{\Ga}_{\lambda\, a}{}^b} \right) \, ,
\ee
where indices and covariant derivatives are all now defined in terms of
the induced metrics $g_{AB}$ and $g_{ab}$ on the corresponding submanifolds. 

Let us stress, already at this point, an important feature of (\ref{finalGa}): 
It was important, for our derivation, to carry out our differentiations at 
fixed $\vphi_*$. In other words, we have been forced to work with
$\Ga_{\la}[\vphi_*, \bar\phi]$. We believe, instead, that no  closed 
RGF-equation holds for the original Vilkovisky-De Witt effective action 
$\Ga_{_{VDW}}[\bar\phi] \equiv \Ga[\bar\phi, \bar\phi]$.
This is probably related to the fact that, unlike $\Ga[\vphi_*, \bar\phi]$, 
$\Ga_{_{VDW}}[\bar\phi]$ does not generate
1PI vertex functions \cite{DeWitt,kunst}. 
Since these vertex functions are related to  operators that depend explicitly 
on $\vphi_*$,  it is not surprising that the same is true for the RGF equation. 
This is indeed apparent 
through the second term appearing in the definition (\ref{barGa}) of  
$\bar{\Ga}_{ij}$. Note that the presence of this term, one of the main 
novelties of our paper, is {\it not} required by reparametrization invariance: 
both terms in (\ref{barGa}) are fine from this point of view. It is required 
instead by $\vphi_*$-dependence and thus, we believe, it is a necessary price 
to pay for the whole procedure to work.

Let us see now how the previous steps can be repeated in the case of a
gauge theory. As it was shown by Vilkovisky and DeWitt
\cite{Vilko,DeWitt}, we first need to reduce the gauge theory 
 to a ``non-gauge" one. Let us indicate as before
by $\cal M$ the field space, by $\phi^i$ the gauge fields,
 with $g_{ij}$ the associated metric, 
by $\sigma^m$ a complete set of
gauge-invariant coordinates,  and by $R^i_\al$
the generators of the gauge transformations:
\be
\delta\phi^i=R^i_\al d\eps^\al \ ,
\ee
where $\eps^\al$ are coordinates on the gauge orbits.
The metric decomposes into the block diagonal form \cite{kunst}
\be
ds^2=h_{mn}d\si^m d\si^n+\ga_{\al\beta}d\eps^\al d\eps^\beta \; , 
\ga_{\al\beta}=R^i_{\al}g_{_{ij}}R^j_{\beta} ,
\ee
where 
\be
h_{{mn}}=\frac{\partial \phi^i}{\partial \si^m}
\frac{\partial\phi^j}{ \partial \si^n}~\Pi_{ij} \ ,
\ee
and we defined the projector on physical orbit space
\be\label{proj}
\Pi_i^j=\delta_i^j -  g_{ik}R^k_{\al}\ga^{_{\al\beta}}R^j_{\beta} \ .
\ee
Although the $\sigma^m$ were so far arbitrary, we used
 an important result
of \cite{Vilko} to take them 
as gaussian normal coordinates both in the induced metric $h_{mn}$
and in the full space (provided geodesics are defined, in the latter, with
 respect to Vilkovisky's connection \cite{Vilko}).

Instead of using $\eps^\al$ to parametrize
points on orbits one can start with the ``gauge fixing'' 
coordinates $\chi^\al$ and write the definition of the effective 
action a la Faddeev-Popov: 
\be\label{gamgau} 
e^{-{\Ga}[\vphi_*, \bar\phi]} =
\int[D\phi^i] \sqrt{g}\delta(\chi^\al)\det
\left(\frac{\pa\chi^\al}{\pa\eps^\beta}\right) 
\ e^{- S(\phi)+ ({\si}^m - \bar\si^m)
(D^{-1})_m^n \Ga,_{n}} \; .
\ee
Changing integration variables to $\sigma^m$, $\eps^\alpha$ we get
\bea
e^{-{\hat\Ga}[\vphi_*, \bar\si]}
&=&\int[D\si^m] [D\eps^\al] \sqrt{h} \sqrt{\ga}\delta(\chi^\al)\det
\left(\frac{\pa\chi^\al}{\pa\eps^\beta}\right) 
e^{- S(\phi_*,\si)+ ({\si}^m - \bar\si^m)
{\hat\Ga},_{m} }\nn\\
&=&\int[D\si^m] \sqrt{h} e^{- \tilde{S}(\phi_*,\si)+ ({\si}^m - \bar\si^m)
{\hat\Ga},_{m} } \; ,
\eea
where
\be\label{tilS}
\tilde{S}=S-\frac12\ln\det(\ga)\ .
\ee
With the gauge 
effective action written in this form we 
 can directly apply the procedure followed from 
 (\ref{gamkhat}) to (\ref{finalhat}) and obtain, as before, 
\be \label{finalhatg} 
{\hat\Ga}_{_{\lambda - \de \lambda}}[\vphi_*, \bar\si] = 
{\hat\Ga}_{_{\lambda}}[\vphi_*, \bar\si]  + \frac12
\ln ~ \left( \frac{{\rm det}~ \hat\Gamma_{\lambda},_{A}{}^{B}}
{{\rm det}~ \hat\Gamma_{\lambda},_{a}{}^b} \right) \, .
\ee
We can now repeat the steps (\ref{defGamma})-(\ref{finalGa}) and,
following \cite{Vilko,DeWitt,FT}, write 
(\ref{finalhatg}) in arbitrary coordinates $\bar\phi$ as
\be \label{finalGag} 
{\Ga}_{_{\lambda - \de \la}}[\vphi_*, \bar\phi] = 
{\Ga}_{_{\lambda}}[\vphi_*, \bar\phi]  + \frac12
\ln \left[ \frac{\det (P_{_{\ge}} \Pi \bar{\Ga} \Pi P_{_{\ge}})}
{\det(P_{_{>}} \Pi
 \bar{\Ga} \Pi P_{_{>}})} \right] \, ,
\ee
where $\bar{\Ga}$ is defined as in (\ref{barGa}) in terms of the
Vilkovisky connection, $\Pi$ stands for the projector on the 
physical space (\ref{proj}), 
and $P_{_{>}}$ ($P_{_{\ge}}$) is a projector 
on ${\cal M_{_{>}}}$ (${\cal M_{_{\ge}}}$). 
Eq.(\ref{finalGag}) is our desired gauge-invariant RG-flow equation for 
$\Ga_{\la}[\vphi_*, \bar\phi]$. As we already stressed in the non-gauge 
case, no  closed RGF-equation is expected to hold for the original VDW 
effective action.

As a check of (\ref{finalGag}) we can compute the one-loop effective 
action and  compare it with \cite{FT}. Within this approximation we 
can set $\Gamma = S$ on the rhs of the definition (\ref{barGa}) of 
$\bar\Gamma$. Noting that $S$ is only a function of $\bar\phi$, we have 
the freedom to set  $\vphi_* = \bar\phi$, yielding 
$\bar\Gamma_{ij} = S_{;ij}$ in (\ref{finalGag}). We finally integrate 
the evolution from $\la=\La$ to $\la=0$. Using $\Ga_{\La} = \tilde S$, 
together with (\ref{tilS}), we get:
\be
\Ga_0  = S+ \frac12\ln  \frac{\det(\Pi_i^k S_{;k}^l \Pi_l^j)}{\det~\ga}
     \, ,
\ee
in agreement  with the one-loop result of \cite{FT}.

Beyond one-loop, our evolution equations 
should be useful in a variety of problems
pertaining to non-abelian gauge theories and to quantum gravity. In practice,  
one will necessarily have to
resort to some form of truncation of $\Ga_k$, so that our exact equations
become approximate RG-flow equations for a finite set of gauge-invariant 
low-energy parameters.  
A potential complication, at this stage, is represented by the explicit 
appearance, in the definition of $\bar\Gamma$, of the base point $\vphi_*$ 
and of the geodesic coordinates built around it. It is not excluded,
however, that this can be turned to one's advantage by a judicious choice 
of $\vphi_*$. Work is now in progress in addressing this kind of questions 
within  specific examples such as non-linear $\sigma$-models, gauge theories, 
and quantum gravity.

We thank M. Bonini, G. Marchesini, G.C. Rossi, K. Yoshida, 
and particularly  G.A. Vilkovisky,  for useful
discussions.  We also wish to acknowledge the support of a ``Chaire
Internationale Blaise Pascal", administered by the ``Fondation de l'Ecole
Normale Sup\'erieure'', during the early stages of this work, at the
Laboratoire de Physique Th\'eorique, Orsay.  K.A.M. thanks the
Theory Division at CERN for hospitality. 
K.A.M. was partially supported by a Polish KBN grant
and the European Programme HPRN--CT--2000--00152.

\end{document}